\documentclass[11pt]{article}

\usepackage{amsfonts}
\usepackage{amsmath}
\usepackage{latexsym}
\usepackage{amssymb}
\usepackage[mathcal]{euscript}
\usepackage{amscd}

\def\ben{\begin{equation}}
\def\een{\end{equation}}
\def\bea{\begin{eqnarray}}
\def\eea{\end{eqnarray}}
\input amssym.def
\input amssym.tex

\begin{document}
\title{ Newton-Hooke spacetimes, Hpp-waves \\
and the cosmological constant}
\author{G.W. Gibbons\footnote{\texttt{g.w.gibbons@damtp.cam.ac.uk}}\phantom{,}  and C.E. Patricot\setcounter{footnote}{6}\footnote{\texttt{c.e.patricot@damtp.cam.ac.uk}} \\
\emph{D.A.M.T.P.,} \vspace{-2mm}\\
\emph{Cambridge University,}  \vspace{-2mm} \\
\emph{Wilberforce Road,}  \vspace{-2mm}\\
\emph{Cambridge CB3 0WA,}  \vspace{-2mm}\\
\emph{U.K.}}
\maketitle



\begin{abstract}
We show explicitly  how the Newton-Hooke groups
$N^\pm_{10}$ 
act as symmetries of  the 
equations of motion of non-relativistic cosmological models
with a cosmological constant. We give the action on the
 associated non-relativistic spacetimes $M^\pm_4$ 
and show how these may be obtained from
a null reduction of  5-dimensional homogeneous pp-wave
Lorentzian spacetimes $M^\pm_5$. This allows us to realize  
the Newton-Hooke groups and their Bargmann type central
extensions as subgroups of the isometry groups of $M^\pm_5$.
The  extended Schr\"odinger type conformal group is identified
and its action
on the equations of motion given. The non-relativistic conformal
symmetries also have applications to time-dependent harmonic
oscillators. Finally we comment 
on a possible  application to Gao's generalization of the matrix model.      
\end{abstract}

\newpage

\section{Introduction}

Recent  observations of the Cosmic Microwave Background
support the idea  that 
the motion of the universe was  dominated by 
a large positive  cosmological term during a period of primordial inflation
in the past. These  and observations of type Ia supernovae
also suggest that the universe is presently entering another phase
of exponential expansion due to a much smaller positive 
value of  cosmological 
constant $\Lambda$. Reconciling these facts 
with fundamental theory such as M or
String theory or indeed with the most elementary notions  of quantum field
theory is not easy. On the other hand, a negative cosmological term
plays an essential
role in the AdS/CFT correspondence and attempts to establish
whether some sort of Holographic Principle holds in Quantum Gravity.

This suggest that we still lack an adequate understanding  
of the basic physics associated  with the cosmological constant
and that it is worthwhile examining it from all possible
angles. Moreover, if the cosmological constant really is  non-zero at
present, then some  processes at least are  going on right now in which 
its  effects are decisive. A standard general  strategem for understanding
any  physical process is to consider a limiting situation
and see whether any simplifications occur, for
example whether the symmetries of the problem
change or possibly become enhanced. In the case of the cosmological
constant
the relevant symmetry groups are the de-Sitter or Anti-de-Sitter
groups which are {\it relativistic symmetries} involving the 
velocity of light $c$. One  possible limit is the non-relativistic
one in which  $c \to \infty$ and $\Lambda \rightarrow 0$ 
but keeping $c^2 \Lambda $ finite, in which the de-Sitter or
Anti-de-Sitter groups become what are called the Newton-Hooke groups \cite{LB},
the analogues of the Galilei group in the presence of  a universal
cosmological repulsion or attraction\footnote{It is important to
take $\Lambda$ to zero as $c$ goes to infinity, otherwise, as we shall
see later, one looses the boost symmetry }.    
It is this limit which we propose studying in this paper.
We shall begin by setting up the basic equations for non-relativistic
cosmology with a cosmological term and then exhibit the action
on the solutions of these equations of the Newton-Hooke groups and their
central and conformal extensions, the analogues of the Bargmann and
Schr\"odinger groups. The Newton-Hooke groups act on a non-relativistic 
spacetime, the analogue of Newton-Cartan spacetime for the Galilei
group but the geometrical structures involved are  complicated.     
The picture simplifies dramatically if one regards this generalized
Newton-Cartan spacetime as a Kaluza-type null reduction of 
a five-dimensional spacetime with a conventional  Lorentzian
structure. In the case of the Galilei group the five-dimensional
spacetime is flat; in the case of the Newton-Hooke groups it turns
out to be a homogeneous pp-wave of the same general type that
have been
at the centre of attention recently in
connection with Penrose limits of the AdS/CFT.

The suggestion that the Newton-Hooke
algebras could have an application to  non-relativistic
cosmology is not new, it goes back to their very beginnings in 
\cite{LB} and it  was developed to some extent in 
\cite{Derome}. More recently it was  revived in \cite{Aldrovandi},  
motivated precisely by  observations of Type Ia
supernovae. Within M/String theory, and with similar motivations,
 Gao has given a modification of the Matrix model
using the Newton-Hooke group \cite{Gao}. It has also been argued
recently
that the Carrollian contraction of the Poincar\'e group
in which $c \downarrow 0$ 
 is relevant to the problem of tachyon condensation
\cite{GHY}.
 Another  possible limit that has been considered is
that of a  very large  cosmological
constant \cite{ABCP}.   

Quite apart from these rather formal considerations, it is possible
that our work may prove useful in the study of large scale structure
since the  equations of motion we study appear there under the
guise of the
Dimitriev-Zel'dovich\footnote{The name is not quite standard.
We have adopted it because no other suitable name is in general
use and these authors appear to have been the first to write them down}
equations \cite{Dim Zel} and we provide a complete account of
their symmetries.    

The plan of the paper is as follows. In section \ref{non-rel cosmo}
we derive the equations of motion of self-gravitating
non-relativistic particles  in a universe with cosmological constant,
and show that the relevant limit, in order to preserve boost
symmetries, is taking $c\to \infty$ and $\Lambda \to 0$ keeping
$\Lambda c^2$ fixed. We relate the equations to the
Dimitriev-Zel'dovich equation, of which we give a
derivation in the Appendix. We present the Newton-Hooke groups
$N_{10}^{\pm}$ in section \ref{N-H
  groups}, and review the initial motivation of L\'evy-Leblond and Bacry
\cite{LB} who realized these groups could be obtained by an 
In\"on\"u-Wigner contraction of the de-Sitter and Anti-de-Sitter
groups. We then define in section \ref{N-H spacetimes} their
four-dimensional cosets $M_4^{\pm}$, or associated Newton-Hooke
spacetimes. We show  that the
action of $N_{10}^{\pm}$ on $M_4^{\pm}$ precisely corresponds to
changes of inertial frames of a non-relativistic cosmological
model with cosmological constant. In other words, the equations of
motion of such  models are left invariant
under the action of the Newton-Hooke groups, in the same
way as Newton's equations are invariant under Galilean transforms. However,
as their Galilean counterpart the Newton-Cartan spacetime, the
Newton-Hooke spacetimes $M_4^{\pm}$ do not admit invariant
metrics. Burdet, Duval, Perrin and K\"unzle \cite{DBKP} gave an
elegant description of 
Newton-Cartan spacetime in terms of Bargmann structures. We give a
similar description  of $M_4^{\pm}$  in
section \ref{Bargmann}, and it turns out the relevant 5-dimensional
manifolds $M_5^{\pm}$ are homogeneous plane-waves. As one expects the
Heisenberg 
isometry groups of the plane-waves correspond to the centrally
extended Newton-Hooke groups $N_{11}^{\pm}$, where the central extension
represents the mass of particles in $M_4^{\pm}$. Whereas $M_5^{-}$ admits a
causal Killing vector, $M_5^{+}$ does not: this difference is
reminiscent of that between their relativistic counterparts Anti-de-Sitter and
de-Sitter space. In section \ref{non-rel conf}, motivated by
\cite{DGH} and \cite{BDP1} we find the Bargmann
conformal groups or extended Schr\"odinger groups of $M_5^{\pm}$. We
show how these 13-dimensional groups act on the cosmological
equations. They send  solutions with a given
cosmological constant and  
gravitational coupling to solutions with the same cosmological
constant but with a possibly time-varying gravitational coupling, in
an analogous way to the Lynden-Bell transformations \cite{Lynden} in
Newtonian theory. The
symmetries we exhibit also have applications in the theory of
time-dependent harmonic oscillators. In section \ref{matrix model} we
explain that Gao's modification of the Matrix model admits
Newton-Hooke symmetries, and  conclude in section
\ref{conclusion}.

\section{Non-relativistic cosmological constant}
\label{non-rel cosmo} 
In this section we shall derive some of the equations
governing a non-relativistic cosmological model
with a cosmological constant. Let us begin by considering
 the case of a single  non-relativistic particle moving 
in a spacetime with a cosmological constant.
The effect of the
cosmological constant is to provide a repulsive
 ($\Lambda >0$)
or attractive ($\Lambda <0$) force proportional to the distance
from an arbitrary centre leading to  the equation of motion 
\ben
{ d^2 {\bf q}  \over dt ^2 } -{c^2\Lambda \over 3 } {\bf q}=0.
\label{eom}\een     
This can be shown for example by looking at the geodesics in de-Sitter
space. The metric inside the cosmological horizon $r<
\sqrt{\Lambda/3}$ is given by 
\ben
ds^2=-c^2(1-{\Lambda r^2 \over 3})dt^2 +{dr^2 \over (1-{\Lambda r^2
    \over 3})} + r^2(d\theta^2 +\sin^2\theta d\phi^2). \label{de Sitter}
\een
The Hamilton-Jacobi function $S(t,r)$ of a particle of mass $m$ with no angular
momentum about the sphere satisfies
\ben
g^{\mu \nu}\partial_{\mu}S\partial_{\nu}S=-m^2c^2. \label{HJ}
\een
Then $\partial_tS$ is conserved and we have
\ben
\partial_tS=E= mc^2 +\epsilon \nonumber 
\een
where $\epsilon$ is the non-relativistic energy of the particle, and
(\ref{HJ}) becomes
\ben
(1-{\Lambda r^2 \over 3})(\partial_rS)^2={(mc^2+\epsilon)^2 \over
  c^2(1-{\Lambda r^2 \over 3})} -m^2c^2. \label{HJ2}
\een
In the low velocity limit, when  $\epsilon \ll mc^2$ this becomes  
\ben
\Big(1-{\Lambda r^2 \over 3}\Big)^2(\partial_rS)^2= m^2c^2{\Lambda r^2
\over 3} +2m\epsilon, \nonumber 
\een
and in the weak field limit $\Lambda r^2 \ll 1$, if we call
$\partial_r S\equiv p$, this yields 
\ben
\epsilon={p^2 \over 2m}-{m\over 2}{\Lambda c^2\over 3}r^2. \label{HJ3}
\een
It is important  to notice
that this limit  can also be obtained directly
from (\ref{HJ2}) by simultaneously taking  $c\to \infty$ and $\Lambda
\to 0$ provided, 
\ben
\lim {c^2\Lambda \over 3} = {1\over
  \tau^2}.\label{non-rel limit}
\een
When $\Lambda$ is taken to have the
dimensions $[{\rm Length}]^{-2}$, the non-relativistic limit $c\to \infty$
yields an acceptable non-relativistic
motion only if $\Lambda \to 0$  and (\ref{non-rel limit}) is satisfied. As a
consequence (\ref{HJ3}) remains valid for  large $r\ll 1/\sqrt{\Lambda}$.
From this we readily obtain (\ref{eom}), and also the free particle
limit when $\tau \to \infty$. 

One might ask what would have happened if we had merely taken
$c$ to infinity keeping $\Lambda$ fixed.
The corresponding
theory would not have a 10-dimensional kinematical group. It could
still be isotropic, but there would be no boost symmetry.
In detail, the de-Sitter 
algebras contain the bracket 
relation
\ben
[H,P_i] =  {c^2 \Lambda \over  3 } K_i,
\een
where $H$ generates time translations, $P_i$ space translations
and $K_i$ boosts. If we were to take the limit 
$c \uparrow \infty$ but keep
$\Lambda$ fixed, we would have to delete the boost from the algebra.
On the other hand the translations have the bracket
\ben
[P_i, P_i] = { \Lambda \over 3} \epsilon _{i j k} L_k,
\een
where the $L_i$ generate rotations. The corresponding
cosmological models would be of the same type that were developed
in the late nineteenth century,
before the advent of either Special or General Relativity, such as that of
Schwarzschild \cite{Schwarzschild} in which the geometry of space   
was taken to be of constant curvature and independent of time
(${\Bbb R} {\Bbb P} ^3$ in his case)
or more daringly, in the case hinted at by   Calinon \cite{Calinon}
in which the curvature was allowed to vary  with time.
In such models Galilei invariance is broken by the curvature of
space.  The  limit we take is precisely
that needed to get a consistent $c \to \infty$ In\"on\"u-Wigner
contraction of the de Sitter groups to the Newton-Cartan groups.

We could consider more than one particle, a finite number of point
particles, which not only 
experience the cosmological attraction or repulsion, but also
suffer mutual gravitational attractions. Thus if $m_a$ is the mass
of particle $a$ we have (suspending the summation convention for $a$
and $b$)  
\ben
m_a {d^2 {\bf q}_a \over dt ^2 } -  m_a {c^2 \Lambda \over 3} {\bf q}_a=
\sum_{b\ne a} G m_a m_b {( {\bf q}_b -{\bf q}_a) \over |{\bf q}_a -{\bf
    q}_b | ^3} \label{celestial}.  
\een
$G$ is of course Newton's constant. Again, this can be derived from the
low velocity and weak limits in de-Sitter or Anti-de-Sitter spaces
with point particles.  Equations (\ref{eom}) and (\ref{celestial}) are
in fact particular cases of the Dimitriev-Zel'dovich equation
\cite{Dim Zel} (see also \cite{Peebles}) which determines the
non-relativistic motion
of a group of point particles in an expanding homogeneous isotropic
universe with cosmic time $t$. The proper distances of the particles
to the origin are written ${\bf r}_a(t)=A(t){\bf x}_a(t)$, where
$A(t)$ is the scale factor for an F-R-W universe, and ${\bf x}_a(t)$
are comoving coordinates. The Dimitriev-Zel'dovich equation reads:
\ben
{d \over dt}\Big(A(t)^2 {d{\bf x}_a \over dt} \Big)=
{G \over A(t)} \sum_{b\ne a} m_b {( {\bf x}_b -{\bf x}_a) \over |{\bf
    x}_a -{\bf  x}_b  | ^3}. \label{DZ}
\een
For the readers convenience, 
we give a  derivation  of this equation from Newtonian theory
 in Appendix \ref{Dim
  Zel derivation}. If one lets ${\bf x}=a(t){\bf q}$ then (\ref{DZ}) becomes 
\ben
{d^2 {\bf q}_a \over dt ^2 } -{c^2 \Lambda(t) \over 3} {\bf q}_a=
G\sum_{b\ne a} m_b {( {\bf q}_b -{\bf q}_a) \over |{\bf q}_a -{\bf
    q}_b | ^3} \label{celestial2}  
\een
provided that $a(t)=1/A(t)$, and with 
\ben
\Lambda (t)=-{3 \over c^2a^2}(\ddot{a}a -2 \dot{a}^2)={3\over
  c^2}{\ddot{A} \over A} \nonumber
\een
Thus  (\ref{DZ}) describes what we could call non-relativistic motion
of particles in a universe with 
time-dependent cosmological 'constants' $\Lambda (t)$. In fact, as
the equation 
\ben
\ddot{A} -{c^2\Lambda (t)\over 3}A =0
\een
always has solutions $A(t)$ given $\Lambda (t)$, 
(\ref{celestial2}) and (\ref{DZ}) are equivalent. 
One obtains 
$\Lambda=0$ for $A(t)=A_o + B_0t$, and the linear dependence in the
scale factor is reminiscent of the Galilean invariance of (\ref{DZ})
in this case. For
de-Sitter space with $\Lambda>0$ constant, one recovers the scale
factor $A(t)=A_0\cosh (t/\tau)+ B_0\sinh (t/\tau)$, with $\tau=
\sqrt{3/\Lambda c^2}$ the Hubble time. \\
As we shall see in section \ref{non-rel conf}, the conformal geometry of
Newton-Hooke spacetimes and their associated homogeneous plane-waves,
together with that of certain time-dependent plane-waves,  
provides  a description of the symmetries of (\ref{DZ}) and
(\ref{celestial2}).\\

\section{The Newton-Hooke groups}
\label{N-H groups}
The two Newton-Hooke groups $N_{10} ^{\pm}$  appear to have first surfaced
in the work of L\'evy-Leblond and Bacry \cite{LB} (see also \cite{BN})
who classified the possible ten-dimensional kinematic  Lie algebras.
The commutation relations of their Lie algebras $\frak{n}^\pm_{10}$  are 
\begin{alignat}{3}
&[J_i,J_j]= \epsilon_{ijk} J_k,& \qquad &[J_i, P_j]=\epsilon_{ijk} P_k,&
\qquad &[J_i, K_j]=\epsilon_{ijk} K_k, \nonumber \\
&[P_i, P_j]=0,& \qquad  &[K_i, K_j]= 0,& \qquad &[K_i, P_j]=0,
\nonumber \\ 
&[H, J_i]=0,&\qquad &[H, P_i]=\pm {1 \over \tau ^2}  K_i,& \qquad &[H,
  K_i]=P_i, \label{comm rel}
\end{alignat}
where the latin indices run from 1 to 3.
Thus the $J_i$ generate rotations, the $P_i$ are to be thought of
as generating (commuting)
space translations, and the $K_i$ as generating (commuting)
boosts, which also
commute with space translations. Finally $H$ should be thought of as
generating time translations
and commutes neither with boosts nor with space translations.
One description of the groups $N_{10}^-$ and $N_{10}^+$ is as the
semi-direct products 
$ (SO(3) \times SO(2)) \otimes_L {\Bbb R}^6$ and $(SO(3) \times SO(1,1))
\otimes _L {\Bbb  R}^6$ respectively, 
where  $SO(3) \times SO(2)$ and 
$SO(3) \times SO(1,1) $ are  the subgroups
generated by the $J_i$ and $H$ acting by automorphisms on  the abelian subgroup
${\Bbb R} ^6$ of boosts and translations generated by the $ K_i$ and
$P_i$. \\
L\'evy-Leblond and Bacry  recognized that
 $\frak{n} ^-_{10}$ and  $\frak{n} ^+_{10}$ can be respectively obtained as
In\"on\"u-Wigner contractions \cite{Inonu} of the Anti-de-Sitter algebra  $\frak{so}(3,2)$
 and the de-Sitter algebra $\frak{so}(4,1)$, in a similar
way to how one obtains the Galilei algebra $\frak{gal}(3,1)$ 
 as a contraction
of the Poincar\'e algebra $\frak{e}(3,1)$ 
in the limit that the speed of light
goes to infinity. However one must also rescale the cosmological
 constant in the limit in order to get a finite parameter $\tau$ which
 turns out to be 
\ben
\label{tau}
{ 1 \over \tau^ 2} = \pm {c^2 \Lambda \over 3}.
\een
This is completely analogous to the low velocity and weak field limits
to obtain (\ref{eom}). Thus the Newton-Hooke algebras $\frak{n}^\pm_{10}$ depend upon
the  real parameter $\tau$ (taken positive) which has  the dimensions of time, and when this
is taken to infinity, one obtains the Galilei algebra as is easily
seen looking at the commutation relations above.
Since the Poincar\'e algebra is also a contraction
of the de-Sitter or anti-de-Sitter algebras as $\Lambda \rightarrow
 0$, one obtains
a commutative diagram of group contractions, with $\tau$ defined by
(\ref{tau}),

\ben
\begin{CD}
SO(3,2) @ > {\Lambda \uparrow 0 } >>E(3,1) @<{ \Lambda \downarrow 0}
<< SO(4,1)  \\
@VV{c \uparrow \infty, \Lambda \uparrow 0}V @VV{c\uparrow \infty}V
@VV{c\uparrow  \infty, \Lambda \downarrow 0 }V\\
N_{10} ^{+}  @> {\tau \uparrow \infty} >>  {\rm Gal} (3,1) @<{ \tau
  \uparrow \infty} << N_{10}^-
\end{CD}
\een
One sees that the Newton-Hooke groups are to the de-Sitter groups what the
Galilei group is to the Poincar\'e group, and are to the Galilei group
what the de-Sitter groups are to the Poincar\'e group. Hence, as was
suggested by L\'evy-Leblond and Bacry, they can be thought of as the
kinematical groups 
of non-relativistic cosmological models.  This will be justified in
the next section. Of course one can define the
Newton-Hooke groups for $n+1$ dimensional isotropic spacetimes in the
obvious way,  and
get $ (SO(n) \times SO(2)) \otimes_L {\Bbb R}^{2n}$ and $(SO(n) \times SO(1,1))
\otimes _L {\Bbb  R}^{2n}$.  These groups have the same dimension as the
standard kinematical groups of $n+1$ dimensional maximally symmetric
spacetimes, 
$SO(n,2)$, $SO(n+1,1)$, $E(n,1)$ and
${\rm Gal}(n,1)$. One obtains the same contraction diagram.   \\
L\'evy-Leblond and Bacry also  recognized the connection of
$N^-_{10}$, obtained by contraction of $SO(3,2)$,  with systems
of oscillators. This is already clear from the fact
that for $N^-_{10}$ the Hamiltonian $H$  generates a periodic
time translation of period $2 \pi \tau$: indeed using the
Baker-Campbell-Hausdorff formula one gets
\begin{align*}
\exp(tH) P_i \exp (-tH) &= \cos (t/\tau)P_i - {1 \over \tau  } \sin
(t/\tau) K_i \\ 
\exp (t H) K_i \exp (-tH) &= \cos (t/\tau) K_i + \tau \sin (t/\tau) P_i.
\end{align*}
Similarly in $N^+_{10}$ one gets
\begin{align*}
\exp(tH) P_i \exp (-tH) &= \cosh (t/\tau)P_i + {1 \over \tau  } \sinh
(t/\tau) K_i \\ 
\exp (t H) K_i \exp (-tH) &= \cosh (t/\tau) K_i + \tau \sin (t/\tau) P_i.
\end{align*}
so that the Hamiltonian $H$ generates a motion
which is periodic in imaginary time, which is 
indicative of  physics at non-zero temperature. In fact the way we
have set things up is such that it suffices
to take $\tau$ to $i\tau$ in the formulae concerning $N^-_{10}$ to get
the corresponding formulae for $N^+_{10}$.

\section{Newton-Hooke spacetimes}
\label{N-H spacetimes}
The cosmological interpretation of $N^{\pm}_{10}$ and their connection
with oscillators can be made rather 
more concrete by using the full group composition law obtained 
by exponentiating the entire algebra and
exhibiting its action on the Newton-Hooke spacetimes $M_4^{\pm}
$. These are defined as the four-dimensional coset  spaces $\exp (tH+
q_iP_i)$ of the Newton-Hooke groups, i.e as
the quotients of $N_{10}^{\pm}$ by the subgroup $SO(3)\otimes_L
\mathbb{R}^3$ generated by the $J_i$
and the $K_i$. This is analogous to defining Minskowski space as the
coset $E(3,1)/SO(3,1)$ ; in fact Bacry and Nuyts \cite{BN} have shown that all
10-dimensional kinematical groups have a four dimensional space-time
interpretation. \\
Under the action (by left translation) of a group element 
\ben
(t_0, a_i, v_i,R)=\exp(t_0 H)\exp(a_iP_i)\exp(v_iK_i)R \nonumber
\een
with $ R=\exp (n_iJ_i)\in SO(3)$, the coordinates of a space-time point   
$(t,q_i)$ are  transformed to, in the case of $N_{10}^-$: 
\begin{align}
t &\rightarrow t+t_0, \nonumber \\
q_i &\rightarrow (Rq)_i + v_i \tau \sin (t/\tau) + a_i \cos (t /\tau).        
\label{Hooke1}
\end{align}
In the case of $N^+_{10}$ one must replace the
trigonometric functions by hyperbolic-trigonometric functions, or
$\tau$ by $i\tau$. 
The action of $N_{10}^{\pm}$ on $M_4^{\pm}$ represents a change of
coordinates between two Newton-Hooke 'inertial' frames,
the second being obtained from the first by successively making a
rotation R, a Newton-Hooke boost of velocity $v_i$, a
space translation by $a_i$, and a time translation by $t_0$. The parameters are given in the first frame. Clearly in the limit $\tau \rightarrow \infty$, for both $M_4^+$ and $M_4^-$, 
we obtain the standard formulae
for Galilei transformations acting on Newton-Cartan space-time
$M_4^0$. Note also that $t$ is an absolute time in $M_4^{\pm}$. \\
The Newton-Hooke transformations mathematically define the
changes between 'inertial' frames in the space-times
$M_4^{\pm}$, but we can see now that these
transformations indeed represent what one would define as inertial
transformations of a non-relativistic cosmological model with
cosmological constant. The equation of motion (\ref{eom}) is clearly invariant
under the action of the Newton-Hooke group (\ref{Hooke1}).
Furthermore, it is easy to see that (\ref{celestial}) is also  invariant
under the action (\ref{Hooke1}). In fact we could replace Newton's law
of gravitational attraction 
by any other central force law and still obtain an equation
of motion invariant under the action of the Newton-Hooke group. Thus if $\{{\bf
  q_a}(t) \}$ is a set of trajectories describing the motion of
non-relativistic particles of mass $m_a$ experiencing mutual
gravitational attraction and cosmological
attraction (or repulsion), then so is its image under any
Newton-Hooke transformation. In other words, the Newton-Hooke
coordinate transformations are indeed inertial transformations,
where inertial now refers to the laws of physics described by equations
such as (\ref{celestial}). \\

We have constructed the Newton-Hooke spacetimes $M^\pm_4$
 as the homogeneous spaces
$N_{10}^{\pm}/ (SO(3)\otimes_L \mathbb{R}^3)$;
in fact they are  symmetric spaces. Indeed, if $\frak{h}$ is the Lie
subalgebra of $\frak{n}_{10}^{\pm}$ spanned by the $J_i$ and $K_i$,
and $\frak{q}$ the vector space spanned by the $P_i$ and $H$, then we
have: 
\ben
\frak{n}_{10}^{\pm}=\frak{h}\oplus \frak{q}, \quad [\frak{h},
  \frak{h}]\subset \frak{h}, \quad [\frak{h}, \frak{q}]\subset
\frak{q}, \quad {\rm and} \;\; [\frak{q}, \frak{q}]\subset \frak{h}, \nonumber
\een
so that $\frak{n}_{10}^{\pm}=\frak{h}\oplus \frak{q}$ is a symmetric
split. Since $SO(3)\otimes_L \mathbb{R}^3$ is not a normal subgroup,
$M_4^{\pm}$ are not group manifolds. One can represent the infinitesimal
generators of the group action (\ref{Hooke1}) by
vector fields on $M_4^+$
\begin{alignat}{2}
&H= \partial _t,&  \qquad &P_i= \cos (t /\tau) \partial _i, \nonumber \\
&K_i = \tau \sin (t/ \tau) \partial _i,& \qquad &J_i=\epsilon_{ijk}q^k
\partial_j, \label{Killing} 
\end{alignat}
and correspondingly for $M_4^-$. Newton-Hooke
spacetimes $M_4^{\pm}$, like their 
limit Newton-Cartan spacetime $M^0_4$, do not admit an invariant
non-degenerate metric: indeed, the co-metric given at the origin
by a symmetric tensor 
\ben
L=g_{tt}H^2 + g_{tj}HP^j + g_{jt}P^jH + g_{kl}P^kP^l,\nonumber
\een
needs to be invariant under
the adjoint action of $SO(3)\otimes_L \mathbb{R}^3$ for the metric to be
well-defined on the coset $M_4^{\pm}$. Since $[K^i, L]=0$
implies that $g_{tt}=g_{jt}=0$, $L$ is necessarily a degenerate co-metric,
and one cannot define the assosiated metric. Nevertheless
 $M_4^{\pm}$ do have a well defined geometric structure which 
may be described in a covariant fashion. One description
proceeds by endowing $M_4^\pm$ with degenerate co-metrics (or
contravariant metrics)
and an affine connection. 
The co-metrics can be obtained group theoretically
($L=\delta_{ij}P^iP^j$) or by taking the limits
$c\to \infty$ and $\Lambda\to 0$ of the Anti-de-Sitter and de-Sitter
metrics (see (\ref{de Sitter})).   
However, just as in the case
of Newton-Cartan spacetime, it proves more convenient to 
regard Newton-Hooke spacetimes as a Kaluza type null reduction of an  ordinary
5-dimensional  Lorentzian spacetime
$M_5^{\pm}$. We shall see in the next section that $M_5^{\pm}$ are in fact
homogeneous plane-waves.

\section{Bargmann Structures}
\label{Bargmann}
Duval, Burdet, K\"unzle and Perrin 
\cite{DBKP} have given an elegant  construction 
of Newton-Cartan spacetime $M_4 ^0$  as the  null reduction of 
a certain five-dimensional Lorentzian spacetime $M_5$
 equipped with what they
called a Bargmann structure. This is essentially a covariantly
constant null Killing vector field $V$ generating
an ${\Bbb R}$, or possibly $S^1$, action which we shall call $G_{\rm null}$. 
The idea was further developed
in \cite{DGH} where $M_5$ was shown to be a Brinkmann or pp-wave
spacetime. The merit of the approach is that it exhibits
the action of the Galilei group, its central extension the
 Bargmann group, and the non-relativistic conformal or Schr\"odinger group
 as subgroups of the isometry
group, or conformal group of $M_5$ which commute with the projection
$\pi:  M_5 \rightarrow M^0_4 \equiv M_5/G_{\rm null}$. One 
can apply these  ideas to  Newtonian Cosmology and
the Newtonian N-body problem \cite{DGH}.  

We shall now
adapt this construction to the case of Newton-Hooke spacetimes.    
Consider  the five dimensional plane-wave space-times $M_5^{\pm}$ with
Lorentz metrics    
\ben
\label{p-w metric} ds^2_{\pm} = dq_i dq_i + 2 dt dv \pm { 1 \over \tau ^2 } q_i q_i dt ^2.
\een
In addition to the obvious action of the rotation group $SO(3)$, the
metric  is easily seen to be  invariant under
the eight -parameter group whose action is   
\begin{align}
t &\rightarrow t+t_0, \nonumber\\
q_i &\rightarrow q_i + F_i(t), \nonumber \\
v &\rightarrow v + q_i G_i (t) +F(t), \label{Killing motion}
\end{align}
as long as
\ben
{\dot F}_i + G_{i}=0, \qquad {\dot G}_i  \pm{1 \over \tau^2}
F_i=0,\qquad 2{\dot F} + {\dot F}_i{\dot F}_i \pm {1 \over \tau^2}F_i
F_i=0.\nonumber 
\een
Thus
\ben
{\ddot F}_i  \mp{1 \over \tau^2} F_i =0,\nonumber
\een
and we may we take as its solution in the case of $M_5^{-}$
\ben
F_i= \tau v_i \sin(t/\tau) + a_i \cos (t/\tau), \nonumber
\een
where $v_i$ and $a_i$ are 2 constant 3-vectors (and take $\tau\to i\tau$ for $M_5^{+}$). The function $F(t)$ is determined only up to a constant of integration
$v_0$, so that with $t_0$, the motion (\ref{Killing motion}) has
8 parameters. We explicitly introduce $v_0$ as corresponding to  the  action of  $G_{\rm null}$,
\ben
v \rightarrow v+ v_0,\nonumber
\een  
with associated covariantly constant null Killing vector field 
\ben
V= \partial _v.\nonumber
\een
The Killing vector fields $P_i$ and $ K_i$ of $M_4^{\pm}$ given in
(\ref{Killing}), when lifted to   $M^{\pm}_5$, 
are modified to 
\ben
P_i \rightarrow P_i + 
q_i {1 \over \tau } \sin(t/\tau)  \partial _v, \qquad 
K_i \rightarrow K_i   -q_i \cos(t /\tau) \partial_v,
\een
and their Lie bracket becomes 
\ben
[P_i,K_j]=-\delta_{ij} V. \label{Heisenberg}
\een
The other commutation relations given in (\ref{comm rel}) remain the
same, so that $V$ is a central element. Thus the isometry groups
$N^{\pm}_{11}$ of $M_5^{\pm}$ are non-trivial central extensions of the
Newton-Hooke groups $N_{10}^{\pm}$ by $G_{null}$.     
The informed reader will recognize (\ref{Heisenberg})
as defining the Heisenberg subalgebra of the isometry
algebra of Hpp-waves, with $H=\partial_t$ and $J_i=\epsilon_{ijk}q^k \partial_j$ acting on it as outer
automorphisms. In other words the groups $N^{\pm}_{11}$ have the
structure of the semi-direct products
$(SO(3) \times SO(2)) \otimes _L {\rm H(7)}$ and 
 $(SO(3) \times SO(1,1)) \otimes _L {\rm H(7)}$, where ${\rm H}(7)$ is 
the Heisenberg group generated by $P_i$, $K_i$ and $V$. \\

It is clear
that all the  isometries of $M_5^{\pm}$ commute with
the projection $\pi : M_5^{\pm}\to M_4^{\pm}$ given by
$(v,t,q_i)\mapsto (t,q_i)$. This is
completely analogous to the usual Bargmann group which is the
central extension of the Galilei group, or symmetry group of
Newton-Cartan space-time, and indeed our formulae reduce to that case
in the limit $\tau \to \infty$. Therefore we propose calling
$N^\pm_{11}$ the Bargmann extensions
of the Newton-Hooke groups, or the Bargmann-Newton-Hooke groups for
short. The groups $N^{\pm}_{11} $  have   Casimirs
\begin{align}
c_1&=V,\nonumber \\
c_2& =2VH+P_i^2 \mp{ 1 \over \tau ^2}  K_i^2,\nonumber \\ 
c_4& =(VJ_i-\epsilon_{ijk} K_j P_k )^2.\nonumber 
\end{align}
To pass to the Newton-Hooke group
one sets $V=-m$, where $m$ is the mass of the system
one is considering; $H$ is then its Hamiltonian, $c_2$ its internal
energy and $c_4/m^2$ the square of its spin vector.\\  

It has long been known that while the Anti-de-Sitter group
$SO(3,2)$ admits a well defined notion of positive energy,
no such notion is possible  for the de-Sitter group $SO(4,1)$.
This is connected with the existence of Killing vector fields
on Anti-de-Sitter spacetime $AdS_4$ which are everywhere timelike.
The metric is globally static with respect to these Killing fields
and there are no Killing horizons. By contrast in the case of
de-Sitter spacetime $dS_4$,  there are only locally time-like Killing
fields with Killing horizons beyond which the Killing field
becomes spacelike. In fact the Killing fields generate
$SO(2)$ subgroups of $SO(3,2)$ or $SO(1,1)$ subgroups of $SO(4,1)$
respectively. Note that this is does not mean that de-Sitter space does
not admit a global time function; it does of course. \\
This dichotomy is reflected on the Hpp-wave spacetimes. 
In the case of $M^+_5$ the Killing field $H={\partial \over \partial
  t}= H^{\alpha } {\partial \over \partial x ^\alpha}$
is everywhere causal, becoming  null on the timelike two-surface 
$q_i=0$. By contrast in the case of $M^-_{5}$ the Killing field     
$H={\partial \over \partial t}= H^\alpha  {\partial \over \partial x
 ^\alpha}$ is almost everywhere spacelike, 
becoming null on the timelike two-surface $q_i=0$. Moreover, the
Killing field $K={\partial \over \partial t} -\mu {\partial \over
  \partial v}$ has an ergo-region: it is causal inside the cylinder
$q_iq_i \leq 2 \mu \tau^2$, and becomes spacelike beyond the timelike
hypersurface $q_iq_i= 2\mu \tau^2$. This substancial difference occurs
although both Newton-Hooke spacetimes and their Bargmann manifolds
have an absolute 'time' coordinate, as we see from (\ref{Hooke1})
which remains true in $M_5^{\pm}$. \\
A freely falling   particle of mass $m$ moving on a geodesic in $M^{\pm}_5$ 
has a conserved energy 
$ E=- m g_{\alpha \beta}{d x ^\alpha \over d \sigma}
H^{\beta}$. If the geodesic is future directed and  causal, then in $M^+_5$ 
the energy $E$ can never be negative. By contrast in the case of
$M^-_5 $ the energy can take either sign. Of course when we consider
null geodesics this statement is equivalent to saying that the energy
of a non-relativistic particle in an upside-down potential is
unbounded.

\section{Non-relativistic conformal symmetries}
\label{non-rel conf}
Following the ideas  of Burdet, Perrin and Duval \cite{BDP1}\cite{BDP2} on the
relation between the ''chronoprojective geometry'' of Bargmann
structures and the Schr\"odinger equation, further applied to Newtonian
gravity in \cite{DGH}, we now take a closer look at the conformal
symmetries of $M_5^{\pm}$. As we shall see, the Bargmann structure
(Hpp-wave) introduced to define Newton-Hooke space-times enables us to
find other symmetries of the equations of Newtonian cosmology with
cosmological constant (\ref{eom}) and (\ref{celestial}). \\
Consider a plane-wave space-time with
the following metric in Brinkmann coordinates 
\ben
ds^2= 2dtdv +\alpha(t)q^iq^idt^2 +dq^idq^i. \label{PW}
\een
Since this metric is conformally flat, its conformal
group is locally isomorphic to $SO(5,2)$. As a side remark, this means
that the centrally extended Newton-Hooke groups $N_{11}^{\pm}$,
realized as the isometry groups of $M_5^{\pm}$,  are
subgroups of $SO(5,2)$, in a similar way as is the central extension
of the Galilei group. We find the conformal symmetries of (\ref{PW})
by going to
conformally flat coordinates, which means  going to Rosen coordinates
first, and then redefining $t \to U$ so that the metric scales
properly: let $A(t)$ a solution of
\ben
\ddot{A}-\alpha(t)A=0 \label{diff eq}
\een
and let 
\ben
t=t, \quad v=V -{1 \over 2}\dot{A}(t)A(t)X^iX^i, \quad q^j=A(t)X^j.\label{COC1}
\een
Further define 
\ben
U = \int^{t} {dt' \over A(t')^2}\equiv f(t) \label{COC2}
\een
and the metric (\ref{PW}) becomes 
\ben
ds^2= A(f^{-1}(U))^2(2dUdV + dX^idX^i). \label{flat PW}
\een
For $M_5^-$, we can take $A(t)=\cos(t/\tau)$ and $U=\tau
\tan(t/\tau)$, and for $M_5^+$, $A(t)=\cosh(t/\tau)$ and
$U=\tau \tanh (t/\tau)$. \\
It is easy to see that the null geodesics of (\ref{PW}) can be chosen
to have affine parameter $t$, in which case they satisfy
\ben
\ddot{\bf q}-\alpha(t){\bf q}=0, \qquad \dot{v}=m \label{oscillator}
\een
so that they describe the non-relativistic motion of
particles of mass $m$ in a time-dependent harmonic potential
$-{1\over2}\alpha(t){\bf q}^2$. Since conformally equivalent metrics
have the same null geodesics up to change of coordinates, the null
geodesics of flat space yield all the solutions of (\ref{oscillator});  
however we need to solve (\ref{diff eq}) before hand to find
conformally flat coordinates. We will show in fact that the action of the
conformal group $SO(5,2)$ on (\ref{flat PW}) provides additional symmetries 
 of (\ref{oscillator}) and (\ref{celestial2}). 

Symmetries of time-dependent harmonic oscillators have been
 studied in the past, and extensively used in string theory on time
 dependent plane-wave backgrounds. Lewis and Riesenfeld \cite{Lewis}
 developed a theory of invariants of these systems (time independent
 quantum operators), which Blau and O'Loughlin \cite{Blau} explained
 geometrically in the setting of plane-wave space-times. Essentially
 the isometries  of these spaces (Heisenberg algebra) can be used to
 construct the  invariants. The Bargmann conformal structure yields
 additional symmetries. \\
Burdet, Duval and Perrin define the Bargmann conformal group of a
 pp-metric, or more generally of a Bargmann structure, to be the subgroup
 of the group conformal transformations of the metric which leaves the
 covariantly constant null Killing vector invariant. In other words,
 if $g$ denotes the metric and $\xi$ the Killing vector, a local
 diffeomorphism ${\cal D}$ is a Bargmann conformal transformation if 
\ben
\label{B conf}{\cal D}^*g=\Omega^2g, \quad {\rm and} \quad {\cal D}_*\xi= \xi,
\een
where $\Omega^2$ depends on the coordinates. In our case
$\xi=\partial_v$. As a consequence, a null
 geodesic of (\ref{PW}) with mass parameter $\dot{v}=m$ is mapped to a
 null geodesic with the same mass $m$: the Bargmann conformal symmetries
of (\ref{PW}) leave (\ref{oscillator}) invariant. \\
In practice, ${\cal D}: (t,v,q^i)\to (t^*,v^*, (q^i)^*)$ satisfies
(\ref{B conf}) if and only if ${\cal D}$ is conformal and commutes with the
projection $\Pi: (t,v,q^i)\to (t,q^i)$, and
${\partial v^* \over \partial v}=1$. The conformal
 transformations $(U,V,X^i)\to (U^*, V^*, (X^i)^*)$ of 
\ben
ds^2=dUdV+dX^idX^i \label{flat}
\een
correspond to those of (\ref{PW}). Requiring that the associated tranforms
$t\to t^*$ and  $q^i\to  (q^i)^*$ do not depend on $v$ and ${\partial
  v^* \over \partial 
  v}=1$  is then equivalent to 
\ben
{\partial U^* \over \partial V}={\partial (X^i)^* \over \partial V}=0,
\quad {\partial V^*\over \partial V}=1. \label{B conf 2}
\een 
This follows immediately from the change of coordinates (\ref{COC1})
and (\ref{COC2}). The Bargmann conformal
transformations of (\ref{flat}) are easily found to be \cite{BDP2} \cite{DGH}:
\begin{align}
U^*&= {dU+e \over aU+b} \nonumber \\
V^*&= V +{a\over 2}{(A{\bf X} +{\bf b}U +{\bf c})^2 \over aU+b}
-\langle {\bf b}, A{\bf X} \rangle -{U \over 2}{\bf b}^2 +h, \nonumber\\
{\bf X}^*& = {A{\bf X} +{\bf b}U +{\bf c} \over aU+b}, \label{B conf
  3}
\end{align}
where $A\in SO(3)$, ${\bf b},{\bf c}\in \mathbb{R}^3$; $d,e,a,b,h \in
\mathbb{R}$ with $db-ea = 1$. These transformations define via
coordinate transform the
13-dimensional Bargmann conformal group of the plane-wave metrics
(\ref{PW}). It can be described as the semi-direct product
$(SO(3)\times SL(2, \mathbb{R})) \otimes_L {\rm H}(7)$. The conformal
factor for 
(\ref{flat PW}) is easily seen to be:
\ben
\Omega^2(U)={A(f^{-1}(U^*))^2 \over A(f^{-1}(U))^2}{\partial U^* \over
  \partial U} = {A(f^{-1}(U^*))^2 \over A(f^{-1}(U))^2}{1\over
  (aU+b)^2}.
\een
In terms of the initial coordinates $(t,v,q^i)$, using
$f'(t)=1/A(t)^2$, this becomes:
\ben
\Omega^2(t)= {\partial t^* \over \partial t}, \label{conf factor}
\quad {\rm where} \quad 
t^*= f^{-1}\Big({df(t)+e \over af(t)+b}\Big). 
\een

We now focus on the specific non-relativistic  conformal
transformations given by $A=Id$, ${\bf b}={\bf c}=0$, which generate a
subgroup isomorphic to $SL(2,\mathbb{R})$. We have:
\begin{align}           
t^*&= f^{-1}\Big({df(t)+e \over af(t)+b}\Big), \nonumber \\
{\bf q}^*&= {A(t^*) \over A(t)}{{\bf q} \over (af(t)+b)}. \label{SL2R}
\end{align}
with $db-ea=1$. By construction, these symmetries leave (\ref{oscillator})
invariant. Thus a set of solutions $\{({\bf q}_a(t), m_a ) \}$ of the
time-dependent oscillator equation is mapped
by (\ref{SL2R}) to another set of solutions $\{({\bf q}^*_a(t^*),
m_a)\}$.  When these symmetries have non-trivial conformal factors
(\ref{conf factor}),
they are 'additional': they do not derive from the isometries of
the metric (\ref{PW}). Since the isometry group of (\ref{PW}) is at
most 11-dimensional for $\alpha(t)$ non-trivial, (\ref{SL2R}) does
provide extra symmetries. \\
Consider now a solution $\{({\bf q}_a(t), m_a) \}$ of
Newton's equations  (\ref{celestial2}) with cosmological
'constant' $\Lambda(t)=3\alpha(t)/c^2$ and gravitational coupling
constant $G_o$. We have:
\begin{align}
{d^2 {\bf q}^*_a \over \phantom{(}dt^*\phantom{)}\!\!^2 } -{c^2
  \Lambda(t^*) \over 3} {\bf q}^*_a &=
\bigg( {A(t)\over A(t^*) } (af(t) +b)\bigg)^3\Big(
{d^2 {\bf q}_a \over dt ^2 } -{c^2 \Lambda(t) \over 3} {\bf q}_a\Big)
\nonumber \\
&= {A(t)\over A(t^*) } (af(t) +b) G_o \sum_{b\ne a} m_b {( {\bf q}^*_b
  -{\bf q}^*_a)  \over |{\bf q}^*_a -{\bf
    q}^*_b | ^3},  \label{invariance}  
\end{align}
so that $\{({\bf q}_a(t), m_a) \}$ is taken by (\ref{SL2R}) to a
solution $\{({\bf q}^*_a(t^*), m_a) \}$  with cosmological term 
$\Lambda(t^*)$ but with a time-dependent gravitational constant 
\ben
G(t^*)={A(t) \over A(t^*)}(af(t) + b)G_o. 
\een
When $\Lambda\equiv 0$ (\ref{PW}) is (\ref{flat}), we can take
$A\equiv 1$: this is the case studied in \cite{DGH}, and we recover
the Lynden-Bell symmetries of Newton's equations \cite{Lynden}. 
For $\Lambda<0$ constant, we have
\begin{align}
t^*&=\tau \arctan\Big( {d\tan(t/\tau)+e/\tau \over a\tau
  \tan(t/\tau)+b}\Big) \nonumber  \\
{\bf q}^*&= - \cos(t^*/\tau)\sqrt{(-d+a\tau\tan(t^*/\tau))^2 +{1
  \over \tau^2}(-e +b\tau \tan(t^* /\tau))^2} \;{\bf q}  \nonumber \\
G(t^*)&=-\bigg(\cos(t^*/\tau)\sqrt{(-d+a\tau\tan(t^*/\tau))^2 +{1
  \over \tau^2}(-e +b\tau \tan(t^* /\tau))^2}\bigg)^{-1} 
\end{align}
For $\Lambda>0$, replace $\tan$ and $\arctan$ by $\tanh$ and ${\rm
  arctanh}$. 


\section{The Matrix model}
\label{matrix model}
We recall here the modification proposed by Gao \cite{Gao}
for incorporating de-Sitter physics into the standard matrix model.  
The latter is essentially a non-relativistic
model of a system with non-commuting coordinates. Since it is
formulated in the light-cone gauge 
it  admits  Galilei symmetry, and so Gao, to incorporate
some cosmological features,  constructed a model
admitting  Newton-Hooke symmetry. The equations of motion are
\ben
{ d^2 q_a  \over dt ^2} +{ 1\over \tau ^2} q_a= \sum _b [q_b,[q_a,q_b]],   
\label{matrix} \een
where now the $q_i$ are $N\times N $ hermitian matrices.
The standard matrix model may be obtained by a dimensional reduction
of one spacetime dimension of non-abelian Yang-Mills theory 
with group $U(N)$. One picks a gauge in which the $\frak{u}(N)$
valued connection one form is $A_\mu=(0, q_a(t))$. One could obtain
Gao's modification by breaking gauge invariance by  adding a mass term 
\ben { 1 \over 2 \tau ^2} {\rm Tr } 
A_\mu A^\mu 
\een  
The mass term is``tachyonic'' for $N^+_{10}$, i.e for positive
cosmological constant.  The equations of motion (\ref{matrix}) are
invariant under (\ref{Hooke1}) as 
long as $v_i$ and $a_i$ are multiples of the unit $N\times N$ matrix.
The unit matrices are associated to the centre of mass motion
and are commutative. They do not contribute to the right hand side of 
(\ref{matrix}) and  moreover the Newton-Hooke transformations  
act only on the unit matrices, and so one can say that 
it is a symmetry only of the centre of mass motion, just as 
in the case of self-gravitating particles (\ref{celestial}).   

We have seen earlier how to lift the
 commutative equations (\ref{eom}, \ref{celestial})
to a higher dimensional (mildly) curved spacetime with commutative
 coordinates. This suggests we can use Gao's suggestion to construct a
 higher-dimensional curved spacetime with non-commuting coordinates.    




\section{Conclusion}
\label{conclusion}
We have shown that the Newton-Hooke groups are indeed the relevant
symmetry groups of non-relativistic cosmological models with
cosmological constant. Though we have only considered
Newton's equations in such models, the non-relativistic Schr\"odinger
equation, in its Bargmann formulation \cite{BDP1}, admits
the same symmetries. It is worth noticing that these models, depending
on whether $\Lambda$ is negative or positive, share many common features with
their relativistic counterparts the anti-de-Sitter and de-Sitter cases:
compactness or non-compactness of the 
time generator $H$, definiteness or indefiniteness of the energy of particles,
existence or non-existence of causal Killing fields (not all of which
are independent). It seems therefore that they provide an interesting
geometrical setting for a better understanding of the cosmological
constant.   

Our results, including the action of the  Bargmann conformal group of
$M_5^{\pm}$ on
the cosmological equations of motion, might also  have applications to
string theories  and Matrix
models which have the same symmetries. Moreover, the fact  that
the metric of $M_5^-$ is similar  to that obtained \cite{Blau2} by
taking a Penrose 
limit of $AdS\times S$ spaces raises questions. The former consists of
a non-relativistic limit of $AdS$ lifted to a spacetime with one
extra dimension, while the latter essentially describes the spacetime
(with many non-relativistic features) 
viewed by an observer reaching the speed of light in $AdS\times
S$. In fact, different In\"on\"u-Wigner contractions of the symmetry groups
involved occur, and  yield the same result up to a central extension, the extended Newton-Hooke
group $N_{11}^{-}$.

\appendix
\section{A Derivation of the Dimitriev-Zel'dovich Equations from
  Newton's equations}  
\label{Dim Zel derivation}
We start with the exact equations of motion for a large but finite
number of particles: 
\ben
m_a {\ddot {\bf x}_a}= \sum_{b\ne a} {Gm_am_b
({\bf x}_b -{\bf x}_a) \over |{\bf x}  _a -{\bf x}_b |^3 }
\label{newton2}
\een
and assume that the particles fall into two classes, with
 $a=i, j,k\dots $ and $a=I,J,K,\dots$.
The second set form a cosmological   background and we make the 
approximation that their motion is unaffected by the first  class of
particles, galaxies, 
 whose motion is however  affected both by the background
particles
and their mutual attractions.  
 
Thus the equations of motion (\ref{newton2}) split into two sets
\ben
m_I {\ddot {\bf x}_I}= \sum_{J\ne I} {Gm_I m_J
({\bf x}_J  -{\bf x}_J) \over |{\bf x}  _I -{\bf x}_J |^3 }
\label{background}
\een

and

\ben
m_i {\ddot {\bf x}_i}= \sum_{j\ne i} {Gm_i m_j
({\bf x}_j -{\bf x}_i) \over |{\bf x}  _i -{\bf x}_j |^3 }
 + \sum_{J} {Gm_i m_J
({\bf x}_J -{\bf x}_i) \over |{\bf x}  _i -{\bf x}_J |^3 } \label{galaxies}. 
\een

We now assume that the background particles form  a central configuration

\ben
{\bf x}_I= a(t) {\bf r}_I.
\een
Thus the deviation  of the first set  of particles
from this mean Hubble flow is given by
\ben
m_i {\ddot {\bf x}_i}= \sum_{j \ne i} {Gm_i m_j
({\bf x}_j -{\bf x}_i) \over |{\bf x}  _i -{\bf x}_j |^3 }
 + \sum_{J} {Gm_i m_J
(a(t) {\bf r}_J -{\bf x}_i) \over |{\bf x}  _i -a(t) {\bf r}_J |^3 }
 \label{galxies2}. \een

We replace the absolute positions of the galaxies by the co-moving positions
${\bf x}_i= a(t){\bf r}_i$  and obtain
\ben
m_i\big(a(t){\ddot {\bf r}_i} + 2 {\dot a}(t) {\dot {\bf r}} _i + {\ddot
    a}(t) {\bf r }_i\big)  = {1 \over a^2(t)  } \sum_{j\ne i} {Gm_i m_j
({\bf r}_j -{\bf r}_i) \over |{\bf r}  _i - {\bf r}_j |^3 }
 + {1\over a^2 (t) } \sum_{J} {Gm_i m_J
({\bf r}_J -{\bf r}_i) \over |{\bf r}  _i -{\bf r}_J |^3 } \label{galaxies3}. 
\een
The second term on the right hand side of (\ref{galaxies3}) is the
force ${\bf F}_i$ 
acting on the $i$'th galaxies by the background particles. 
The numerical work in \cite{BGS} provided very good evidence 
that  for a large number of background particles, the
central configuration is to a very good approximation statistically 
spherically symmetric and homogeneous. It follows  that
the force exerted by the background is radial 
\ben  
\sum_{J} {Gm_i m_J
({\bf r}_J -{\bf r}_i) \over |{\bf r}  _i -{\bf r}_J |^3 } = -c m_i{\bf
r} _i,
\een 
where the constant $c$ and  scale factor $a(t)$ satisfy
\ben
a^2 {\ddot a} = -c 
\een  
 
It follows that the force ${\bf F}_i$ on the right hand side of
(\ref{galaxies3}) cancels the third term on the left hand side.  
We are left with 
\ben
m_i \big(a(t){\ddot {\bf r}_i} + 2 {\dot a}(t) {\dot {\bf r}} _i \big)  = {1 \over a^2(t)  } \sum_{j\ne i} {Gm_i m_j
({\bf r}_j -{\bf r}_i) \over |{\bf r}  _i - {\bf r}_j |^3 } \label{DZ2}, 
\een
or
\ben
{ d \bigl (a(t) ^2 { \dot {\bf r}} _i \bigr )  \over dt}
= { 1 \over a(t) }  \sum_{j\ne i} {G m_j
({\bf r}_j -{\bf r}_i) \over |{\bf r}  _i - {\bf r}_j |^3 },
\een
which is the Dimitriev-Zel'dovich equation.

\end{document}